\documentclass{article}

\usepackage{arxiv}

\usepackage[utf8]{inputenc} 
\usepackage[T1]{fontenc}    
\usepackage{hyperref}       
\usepackage{url}            
\usepackage{booktabs}       
\usepackage{amsfonts}       
\usepackage{nicefrac}       
\usepackage{microtype}      
\usepackage{lipsum}		
\usepackage{graphicx}
\usepackage{doi}
\usepackage{amssymb}
\usepackage{listings}
\usepackage{xcolor}
\usepackage{csquotes}

\title{mango: A Modular Python-Based Agent Simulation Framework}

\author{ \href{https://orcid.org/0000-0001-5339-6553}{\includegraphics[scale=0.06]{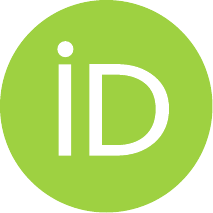}\hspace{1mm}Rico Schrage} \\
	Digitalized Energy Systems Group\\
    Carl von Ossietzky University of Oldenburg\\
    26129 Oldenburg, Germany\\
	\texttt{rico.schrage@uol.de} \\
	\And
	\href{https://orcid.org/0000-0001-6352-4213}{\includegraphics[scale=0.06]{orcid.pdf}\hspace{1mm}Jens Sager} \\
    Energy Division\\
	OFFIS, Institute for Information Technology\\
	26121 Oldenburg, Germany\\
	\And
	\href{https://orcid.org/0009-0008-4535-6218}{\includegraphics[scale=0.06]{orcid.pdf}\hspace{1mm}Jan Philipp Hörding} \\
    Energy Division\\
	OFFIS, Institute for Information Technology\\
	26121 Oldenburg, Germany\\
	\And
	\href{https://orcid.org/0009-0005-6965-1877}{\includegraphics[scale=0.06]{orcid.pdf}\hspace{1mm}Stefanie Holly} \\
    Energy Division\\
	OFFIS, Institute for Information Technology\\
	26121 Oldenburg, Germany\\
}

\renewcommand{\shorttitle}{mango Framework}

\hypersetup{
pdftitle={A template for the arxiv style},
pdfsubject={q-bio.NC, q-bio.QM},
pdfauthor={David S.~Hippocampus, Elias D.~Striatum},
pdfkeywords={First keyword, Second keyword, More},
}

\begin{document}
\maketitle

\begin{abstract}
Agent-based simulations, especially those including communication, are complex to model and execute. To help researchers deal with this complexity and to encourage modular and maintainable research software, the Python-based framework \textit{mango} (modular python agent framework) has been developed. 
The framework enables users to quickly implement software agents with different communication protocols (e.g., TCP) and message codecs (e.g., JSON). Furthermore, \textit{mango} provides various options for developing an integrated agent simulation. This includes a scheduler module, which can control the agents' tasks, a (distributed) clock mechanism for time synchronization, and a specific simulation component, which can be coupled with other (co-)simulation software. These features are complemented by modular implementation patterns and a well-evaluated performance with the ability to simulate across multiple processes to ensure scalability.
\end{abstract}


\begin{table}[!h]
\centering
\begin{tabular}{lp{6.5cm}p{6.5cm}}
\toprule
\textbf{Nr.} & \textbf{Code metadata description} & \textbf{Please fill in this column} \\
\midrule
C1 & Current code version & v1.1.2 \\
C2 & Permanent link to code/repository used for this code version & https://github.com/OFFIS-DAI/mango \\
C3  & Permanent link to Reproducible Capsule & -- \\
C4 & Legal Code License   & MIT \\
C5 & Code versioning system used & git \\
C6 & Software code languages, tools, and services used & Python, protobuf \\
C7 & Compilation requirements, operating environments \& dependencies & Python, paho-mqtt, python-dateutil, dill, msgspec, protobuf \\
C8 & If available Link to developer documentation/manual & https://mango-agents.readthedocs.io/ \\
C9 & Support email for questions & mango@offis.de \\
\bottomrule
\end{tabular}
\caption{Code metadata}
\label{tab:code_meta_data} 
\end{table}

\section*{Introduction}
Complex technical systems are everywhere. Especially today's infrastructure gains in complexity constantly. Most of these systems include many heterogeneous components with non-linear dynamics that interact with each other. The challenge for research software development is to model and simulate these systems to gain insight into their behavior.

One option to deal with these complexities is using agent-based modeling and simulation. An agent, as defined by \cite{russell2010artificial}, is software that perceives its environment through sensors and acts upon that environment through actuators. Designing agents and their behaviors, interactions, and knowledge models is complex. First, it is challenging to develop the different agents, their behaviors, responsibilities, and communication protocols due to the usually complex environment in which they act. Second, implementing these agents to be usable in different use cases while still being maintainable must be rigorously supported by the used agent framework. 

It is difficult to adapt such agents to a research software simulation for large-scale models, agent-based control, and optimization applications while being able to validate the agents in real-world environments. This requires bridging the gap between research and industry while maintaining applicability in both areas. Thus, the framework needs to scale well in a scientific computing environment while running on low-end hardware with network communication. It is also necessary that the agents' implementations for both environments must be identical.

We present \textit{mango} as a new Python-based agent framework that tackles the outlined challenges. It is developed as an evolution of \textit{aiomas}\footnote{https://gitlab.com/sscherfke/aiomas} and is mainly applied in the energy domain. The framework enables simulations with agents running on one or more processes on the same machine. Distributing these agents over different machines is possible without changing their implementations. Furthermore, we propose a novel composition-based role system to support modularization and reusability. On top, the \textit{mango} library contains a collection of common agent functionalities, optimization schemes, and other valuable tools. 

This paper is structured as follows. First, the related work is introduced and discussed. Then, the main functionalities and the architecture of \textit{mango} are described. Furthermore, two applications demonstrate \textit{mango} as a framework. This is followed by a list of projects and publications where \textit{mango} is used, and finally, a short conclusion is drawn.
\section*{Related frameworks}
There are several existing agent frameworks. However, most are no longer maintained, developed for different purposes, or very application-specific. We categorize these frameworks into three different groups.
\paragraph{Agent-based modeling framework} These frameworks focus on building agents in a defined space to model interactions and observe emergent behavior of the complex system. With this approach, it is possible to model various scenarios, such as pandemics, animal behavior, or humans interacting as individuals within their environment. Popular implementations of this type are \textit{mesa} \cite{python-mesa-2020} for Python, \textit{Agents.jl} \cite{Agents.jl} for Julia, or \textit{NetLogo} \cite{tisue2004netlogo}, which even provides a custom language and GUI for the agent modeling. There are two core differences to our presented framework. First, \textit{mango} emphasizes agent communication, including all necessary details like the protocol or the codec. This allows \textit{mango} agents to be usable in real-world applications for validation purposes. Second, while \textit{mango} focuses more on communication, it focuses less on simulation in a predefined (physical) space. Implementing and simulating this in \textit{mango} would also be possible but requires much more work than in these specialized frameworks.
\paragraph{Application-specific agent framework} The term agent framework has multiple meanings, one of which is the application-specific agent framework. These frameworks do not aim to provide a general-purpose tool for simulating agents but provide particular applications in which agents (or agent-like entities) can be created, specialized, simulated, or deployed. For example, \textit{agentMET4FOF} \cite{Met4FoFAgentMET4FOFV0} provides specific tools for meteorological agent-based simulations and can be considered a domain-specific extension to an agent-based modeling framework. The \textit{Wallet Framework for .NET} \cite{AgentFrameworkNET} emphasizes building digital identity wallets based on sovereign identities using the distributed ledger. For this, agents are used as the primary modeling entity. We conclude that \textit{mango} has a different scope, provides more general tools, and is not domain-specific and, therefore, is rather a general-purpose framework.
\paragraph{Multi-agent system framework} The third category emphasizes agent communication, deployability, and a generic approach to modeling agents and their environments. One popular candidate is \textit{JADE} \cite{JADEAgentFramework}, which provides various modeling options to design an agent and its communication. It uses FIPA ACL \cite{o1998fipa} messages for communication purposes. However, \textit{JADE} is not actively developed anymore. It is missing a modular architecture and is developed in Java, which is not popular for scientific applications. Another Java framework is \textit{JIAC} \cite{lutzenberger2013jiac}, which is based on the principle of inversion of control built on the Spring Framework \cite{johnson2004spring}. It has a lot of options to distribute agents on nodes, it is possible to reuse agents' actions, and it has an integrated rule engine to define agent behavior easily. However, it is not maintained anymore and is not applicable outside Java Spring applications. The framework \textit{JACK} \cite{winikoff2005jack} provides a dedicated language to implement agents and tools to build upon the BDI-architecture \cite{broersen2001boid}, but it is also discontinued and proprietary software. In JavaScript, the \textit{agentframework} \cite{agentframeworkNodeJS}, similar to \textit{JIAC}, is based on dependency injection. Lastly, the Python framework \textit{SPADE} \cite{spade} is based on the XMPP platform and does not provide different communication protocols.

We conclude the three categories are not quite comparable, and \textit{mango} is part of the third category. In this category, the frameworks are proprietary, discontinued, or restricted to different areas of agent-based simulation software.
\section*{Software description}
The main objective of \textit{mango} is to create, set up, and execute simulations with communicating agents. Applications can involve distributed control of large systems (e.g., electrical energy systems, traffic lights, etc.), distributed optimization algorithms, and generally any distributed algorithms and behavior analyses. The framework is written in Python with \textit{asyncio}\footnote{https://docs.python.org/3/library/asyncio.html} as a basis to enable asynchronous message exchange in single process applications and to avoid blocking IO in general.

In \textit{mango}, agents can be implemented using an abstract base class or by composition with the help of a role system. Agents \enquote{live} in so-called containers, representing the agents' environment as their gateway to communicate with other agents. The type of the container specifies the communication protocol and the message codec. Messages can be wrapped in FIPA ACL messages \cite{o1998fipa}. On top of the core \textit{mango}, a general-purpose library \textit{mango-library}\footnote{https://github.com/OFFIS-DAI/mango-library} contains implementations for frequent applications like coalition formation.
\begin{figure}
    \centering
    \includegraphics[width=\textwidth]{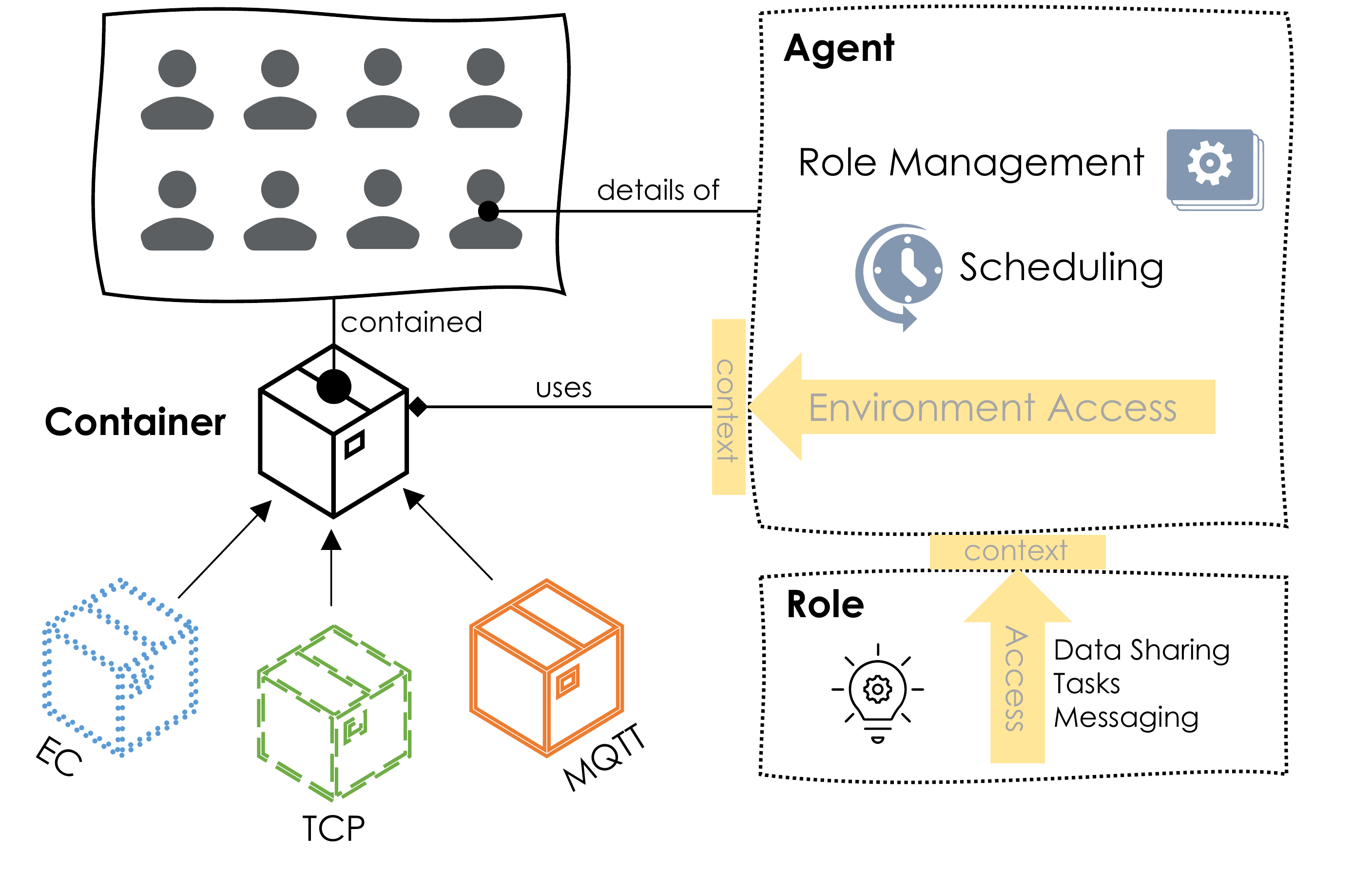}
    \caption{Architecture of \textit{mango}}
    \label{fig:mango_architecture}
\end{figure}
\subsection*{Architecture}
One of the main benefits of \textit{mango} is its modular architecture, depicted in Figure \ref{fig:mango_architecture}, which ensures that most main features can be extended easily. The containers are implemented using a base container class, which implements all common logic. Codecs are designed as plug-ins for containers. New codecs can be implemented by inheriting from \texttt{mango.messages.codecs.Codec} and assigned to the appropriate containers. Every agent inherits the base class \texttt{mango.Agent}. To avoid a cyclic dependency, an abstraction layer is introduced between an agent and its container and between a role and its agent.
Another important aspect is the architectural impact of \textit{mango}. Generally, \textit{mango} does not strictly enforce specific agent architectures due to the different styles of implementing agents and its general loosely coupled, modular structure. However, our philosophy is that a framework should support users to structure their applications to achieve their quality goals. Roles force the user to think about the different responsibilities an agent should fulfill. When used correctly, this role metaphor leads to a strong separation of concerns and supports further modularization due to the more declarative style of implementing the agent logic.
\subsection*{Main functionalities}
In the following, we describe the main functionalities of \textit{mango}. We omit technical code-snipped-aided documentation at this point. We refer to the developer documentation linked in Table \ref{tab:code_meta_data}.
\paragraph{Container} The container is responsible for everything network related to the agent. This includes sending and receiving messages, but also message distribution to the correct agent and (de-)serialization of messages. Messages between agents of the same container are passed directly without creating network traffic.

There are three container implementations: \textit{TCP}, \textit{MQTT}, and \textit{EC}. The TCP container constantly listens to a TCP port and sends all messages using a plain TCP connection. For better efficiency, the connections are buffered and pooled internally. The MQTT container starts an MQTT server using \textit{paho-mqtt}\footnote{https://eclipse.dev/paho/}. The agents can subscribe and send messages to MQTT channels. The last one, the \textit{EC} (external connection) container, can be used to couple \textit{mango} with external simulators or co-simulation tools. This is necessary to set up complex communication simulations with simulators like OMNeT++ \cite{varga2010omnet}.

\paragraph{Agents and roles} The two ways to define agents represent two different, often controversially discussed, approaches to programming object-oriented languages: inheritance and composition. It is not the responsibility of the framework to enforce one of the two approaches but to offer a choice. 

The abstract class \texttt{mango.Agent} for inheritance provides a scheduler and the context object for accessing the container. In most cases, the abstract method \texttt{handle\_message} will be overridden to implement the agents' behavior for specific messages. The method gets called every time a message arrives. For proactive tasks and for sending messages, the scheduler can be used.

For composition, the role system has been implemented. This system uses the generic \texttt{mango.RoleAgent} as the base for adding roles. A role represents one responsibility of the agent as an entity. For example, in the coalition formation, the \texttt{CoalitionParticipant} could be one role whose responsibility is to manage its participation in coalitions. To implement roles, the \texttt{mango.Role} interface can be used. To handle messages, roles can add message handler methods through their role context, representing the role's environment (the agent and the container). Proactive actions are defined using the scheduler. A role can implement lifecycle hook-ins to execute custom logic when created or deleted. There are managed singleton models for which roles can register as observers to share data between roles. Alternatively, there is the option to use a shared data container. 
\paragraph{Codecs} There are two codecs currently implemented: JSON and protobuf\footnote{https://protobuf.dev/}. The codec is defined when creating the container and will then be used for every message exchanged by this container (incoming and outgoing). Both codecs are extendable for custom data classes. The JSON codec supports several Python data structures out of the box. 
\paragraph{Messages} In \textit{mango}, every object can be a message as long it is encodable and decodable by the codec. The FIPA ACL message \cite{o1998fipa} is supported and optionally acts as a wrapper for any custom message.
\paragraph{Scheduling} To realize agent simulations, every agent has a \texttt{scheduler}, which manages all tasks (including tasks for sending messages). The scheduler can start or stop a task and recognize whether a task has been finished or is sleeping. 
\paragraph{Multiprocessing} To ensure the utilization of multi-core systems, the framework uses multiprocessing (due to the lack of real multithreading in CPython) for parallelization. There are two different ways to shift workload to another process. First, the scheduler can execute \texttt{ProcessTasks}. Second, using agent processes will move agents to different processes but link them by IPC to the main container using a mirror container. The first approach should be applied when single long-running tasks must be processed, and the second when agents generally need a lot of time to process and send messages.
\section*{Application}
As \textit{mango} is a general-purpose agent framework, it can be used in various domains without restrictions. Until now, it has mainly been applied to energy domain scenarios. In the following, we introduce a simple example to show how the framework can be used. After that, a more sophisticated example will show an energy case study from a recent research project.
\subsection*{Hello world example}
This example includes the definition of two agents: the Caller-Agent and the Receiver-Agent. The Caller-Agent starts the conversation by scheduling a message for five seconds in the future (see Figure \ref{fig:se_caller}). The Receiver-Agent will answer using the same content (see Figure \ref{fig:se_receiver}). After that, the caller will answer again until 100 messages are sent. This results in a message ping-pong. The simulation is started using a TCP container with a clock, which can be set to a custom time to start the Caller-Agents' initial message. To wait for the results, \texttt{tasks\_complete\_or\_sleeping} is called (see Figure \ref{fig:se_init}).
\lstdefinelanguage{Pythonex}{%
  language     = Python,
  morekeywords = {async, await},
}

\begin{figure}
    \centering
\begin{lstlisting}[language=Pythonex]
class Caller(Agent):
    def __init__(self, container, addr, rid):
        super().__init__(container)
        
        self.schedule_timestamp_task(
            coroutine=self.send_hello_world(addr, rid),
            timestamp=self._context.current_timestamp + 5,
        )
        self.i = 0

    async def send_hello_world(self, addr, rid):
        # send_acl_message to (addr, rid)
    def handle_message(self, content, meta):
        # answer until i == 100
\end{lstlisting}
    \caption{Simple Example: Caller Agent}
    \label{fig:se_caller}
\end{figure}
\begin{figure}
    \centering
\begin{lstlisting}[language=Pythonex]
class Receiver(Agent):
    ...
    def handle_message(self, content, meta):
        self.schedule_instant_acl_message(
            receiver_addr=self._context.addr, 
            receiver_id="agent1", 
            content=content
        )
\end{lstlisting}
    \caption{Simple Example: Receiver Agent}
    \label{fig:se_receiver}
\end{figure}
\begin{figure}
    \centering
\begin{lstlisting}[language=Pythonex]
c = await create_container(addr=addr, clock=clock)
receiver = Receiver(c)
caller = Caller(c, addr, receiver.aid)
await asyncio.sleep(1)
clock.set_time(clock.time + 5)

# wait until each agent is done with all tasks
await receiver._scheduler.tasks_complete_or_sleeping()
await caller._scheduler.tasks_complete_or_sleeping()
\end{lstlisting}
    \caption{Simple Example: Initialization}
    \label{fig:se_init}
\end{figure}
\subsection*{Grid congestion example}
\begin{figure}
    \centering
    \includegraphics[width=\textwidth]{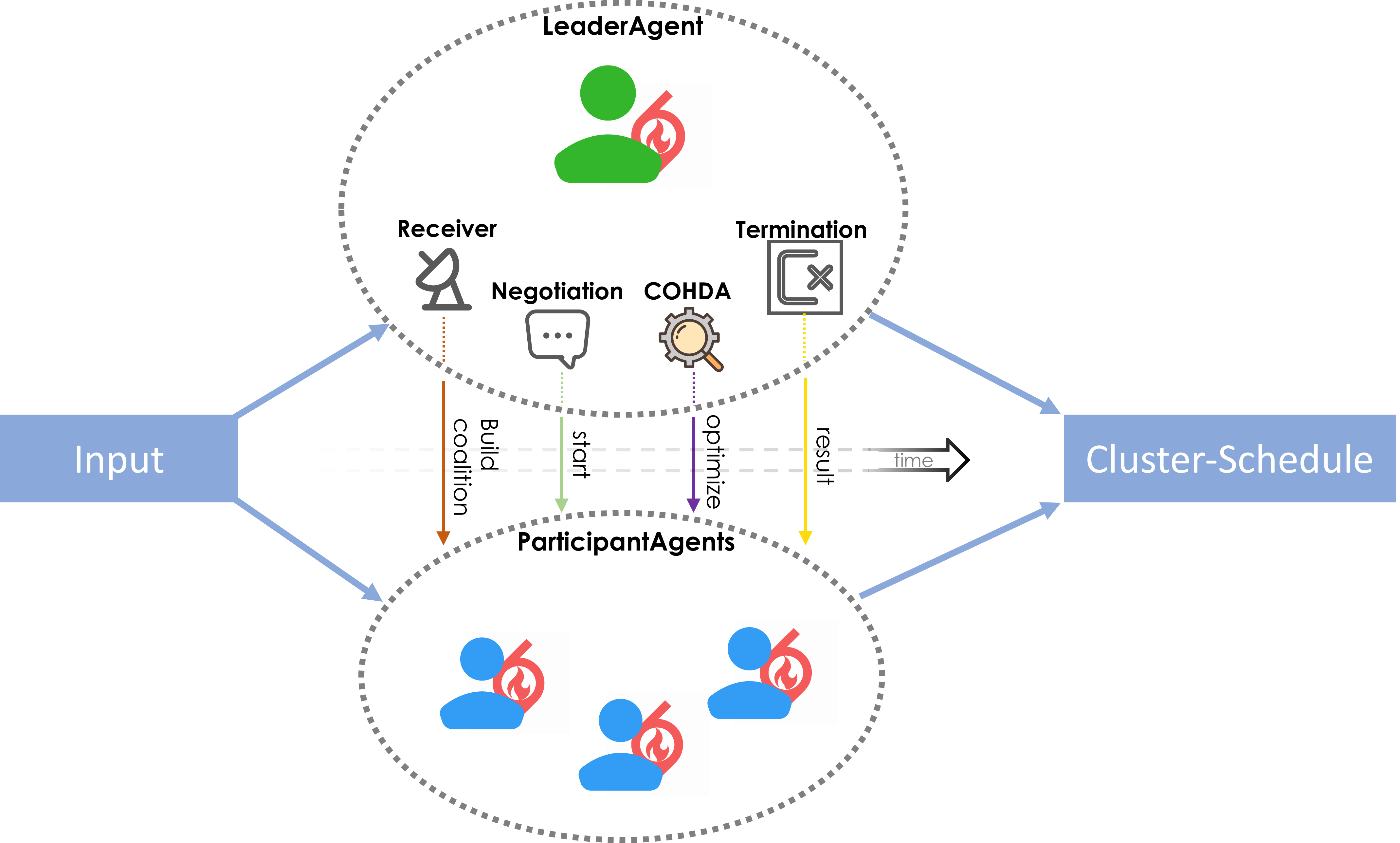}
    \caption{Congestion Scenario: Overview}
    \label{fig:congestion_example_overview}
\end{figure}
In this case study we present a more complex application scenario of the mango framework. It originates from the ZDIN-ZLE\footnote{https://www.zdin.de/zukunftslabore/energie} and is called \textit{mosaik-cohda}\footnote{https://gitlab.com/zdin-zle/models/mosaik-cohda}. 
In power systems, grid operators must ensure that equipment such as transformers or power lines are not overloaded. In this energy scenario, the goal is to prevent an imminent congestion by using the flexibility of multiple heat pumps to decrease energy consumption in a specific area of the grid. 
The heat pumps are controlled by \textit{mango}-agents.

The scenario can be divided into three steps. In the first step, one agent acts as a leader and informs all agents about the objective (a joint reduction of the load profile), thus building the coalition to avoid congestion. 
The second step is to negotiate which heat pump will reduce its load at what time and by how much. 
The agents thus solve a distributed optimization problem. Each agent can deviate from its original schedule through the flexibility of its heat pump. Depending on the load reduction that the other agents can contribute, different own load adjustments are best suited to achieve the desired total load reduction in all time intervals via the cluster schedule. 

This optimization problem is solved using the COHDA heuristic \cite{hinrichs2017distributed}, which is also initiated by the leader agent. Once a cluster schedule is determined, the agents execute it in the simulation by adapting their control commands to the heat pump simulators.

This procedure is embedded in a larger scenario (connected by \textit{mosaik} \cite{app9050923}), in which the grid operator detects the congestion and requests flexibility contribution from specific heat pump agents, so it needs to be executed multiple times. It is also necessary to modularize and reuse some of the existing implementations, as in the same context, numerous other scenarios involve similar objectives but with a different composition (e.g., different flexibility models, different negotiation initiators, etc.). 

The scenario has been implemented using the role system to address these requirements. Therefore, the responsibilities of the agents were identified in the system, and four different roles were built. %
\begin{description}
    \item[FlexReceiverRole] This role is responsible for receiving the flexibility of the agents' heat pump. The flexibility is sent as part of the input values.
    \item[FlexNegotiationStarterRole] This role is responsible for leading the coalition and, therefore, also for starting the negotiation.
    \item[FlexCohdaRole] This role extends COHDA with a schedule generation approach \cite{schrage2023multi} to estimate the optimal flexibility contribution.
    \item[FlexTerminationRole] This role will detect and manage the global termination of the distributed optimization. Consequently, it will also announce the termination by sending the final result.
\end{description}
These roles are complemented by the roles from the \textit{mango-library} to build up the coalition. The process with the participating roles at every step is depicted in Figure \ref{fig:congestion_example_overview}.
\section*{Impact}

The framework is currently actively used for several research projects. In the following, we list the most relevant example publications and projects \textit{mango} is used for.
\begin{description}
\item[Agent Dynamics in Energy Systems] In \cite{Schrage.2023}, the authors used \textit{mango} to model and simulate a multi-agent system, in which agents perform coalition formation based on the energy balance in the electricity, heat, and gas networks. The resulting dynamics were used to analyze the effect of the topology on the coalition formation.
\item[Multi-Purpose Battery Storage Swarm] 
The MIRAGE project\footnote{https://www.offis.de/en/offis/project/mirage.html} was concerned with developing a battery storage swarm based on \textit{mango}. Individual \textit{mango} agents locally optimized battery storages, i.e., generated demand or generation forecasts of other assets, and determined the flexibility of the battery storage according to the fulfillment of their local application \cite{tiemann2022operational}. In aggregate, the flexibility of the swarm was offered to energy markets to generate additional benefits. The \textit{mango} agents could compensate for deviations due to forecast inaccuracies in a self-organized manner. A hybrid field-laboratory test of the \textit{mango} framework was conducted. 
\item[Distributed Blackout Restart]
In \cite{stark2021your}, Stark et al. presented a flexible black start service for restoring an electrical system alongside impaired ICT after a blackout event. The multi-agent system - implemented in \textit{mango} - addressed the restoration of the power system as a distributed optimization problem.
\item[Effect of communication topologies]
In \cite{holly2021dynamic}, the authors employed \textit{mango} to investigate the impact of communication topologies, which provide an overlay that directs communication between agents in distributed optimization heuristics. They also presented an approach for adapting the communication topology at runtime and studied the effects of various static and dynamic topologies on optimization performance in terms of solution quality, convergence speed, and collaboration cost.
\item[Evaluation of ancillary services markets] 
The REMARK\footnote{https://www.offis.de/en/offis/project/remark.html} project develops an open-source toolbox\footnote{https://gitlab.com/remark1/remark-toolbox} for the evaluation of market designs for ancillary services (e.g. reactive power or redispatch). Key actors such as the grid operator, market operators, and market participants are modeled as \textit{mango} agents. Special emphasis is put on the partially AI-based strategic market participants and the investigation of the effects of the market design on the power system. 
\end{description}

\section*{Conclusion}
This paper presents \textit{mango}, a Python-based agent framework that addresses the complexities of agent-based simulations, especially in communication. It is designed for modularity and maintainability and enables quick agent implementation using three communication protocols and two message codecs. The frameworks' real-world relevance and versatility are highlighted through a detailed application to grid congestion and other usages in papers and industry projects. We conclude that \textit{mango} is a valuable tool for bridging the gap between research and real-world scenarios and promotes well-established best practices in software development.
\section*{Acknowledgements}
This work has been partly funded by the Deutsche Forschungsgemeinschaft (DFG, German Research Foundation) – 359941476.

The impulse and basis for the development of the framework originated in the research cooperation with be.storaged GmbH. 

Further development was partly funded by the Lower Saxony Ministry of Science and Culture under grant number 11-76251-13-3/19 – ZN3488 (ZLE) within the Lower Saxony “Vorab“ of the Volkswagen Foundation and was supported by the Center for Digital Innovations (ZDIN).

We thank Marvin Nebel-Wenner for his significant contribution to developing \textit{mango} during his employment at OFFIS.

\bibliographystyle{ieeetr}
\bibliography{main}  

\begin{thebibliography}{10}

\bibitem{russell2010artificial}
S.~J. Russell and P.~Norvig, {\em Artificial intelligence a modern approach}.
\newblock London, 2010.

\bibitem{python-mesa-2020}
J.~Kazil, D.~Masad, and A.~Crooks, ``Utilizing python for agent-based modeling: The mesa framework,'' in {\em Social, Cultural, and Behavioral Modeling} (R.~Thomson, H.~Bisgin, C.~Dancy, A.~Hyder, and M.~Hussain, eds.), (Cham), pp.~308--317, Springer International Publishing, 2020.

\bibitem{Agents.jl}
G.~Datseris, A.~R. Vahdati, and T.~C. DuBois, ``Agents.jl: a performant and feature-full agent-based modeling software of minimal code complexity,'' {\em {SIMULATION}}, Jan. 2022.

\bibitem{tisue2004netlogo}
S.~Tisue and U.~Wilensky, ``Netlogo: A simple environment for modeling complexity,'' in {\em International conference on complex systems}, vol.~21, pp.~16--21, Citeseer, 2004.

\bibitem{Met4FoFAgentMET4FOFV0}
B.~Ludwig, bangxiangyong, A.~Prasad, H.~Lulic, M.~Gruber, and gertjan123, ``{{Met4FoF}}/{{agentMET4FOF}}: V0.13.2,'' {\em zenodo}, 2022.

\bibitem{AgentFrameworkNET}
S.~Bickerle, C.~Hempel, and K.~Dinh, ``Wallet framework for .net (1.6.4), former agent framework for .net [last access 13-11-2023],'' 2023.

\bibitem{JADEAgentFramework}
F.~Bellifemine, A.~Poggi, and G.~Rimassa, ``Jade: A fipa2000 compliant agent development environment,'' in {\em Proceedings of the Fifth International Conference on Autonomous Agents}, AGENTS '01, (New York, NY, USA), p.~216–217, Association for Computing Machinery, 2001.

\bibitem{o1998fipa}
P.~D. O'Brien and R.~C. Nicol, ``Fipa—towards a standard for software agents,'' {\em BT Technology Journal}, vol.~16, pp.~51--59, 1998.

\bibitem{lutzenberger2013jiac}
M.~L{\"u}tzenberger, T.~K{\"u}ster, T.~Konnerth, A.~Thiele, N.~Masuch, A.~He{\ss}ler, J.~Keiser, M.~Burkhardt, S.~Kaiser, and S.~Albayrak, ``Jiac v: A mas framework for industrial applications,'' in {\em Proceedings of the 2013 international conference on Autonomous agents and multi-agent systems}, pp.~1189--1190, 2013.

\bibitem{johnson2004spring}
R.~Johnson, J.~Hoeller, K.~Donald, C.~Sampaleanu, R.~Harrop, T.~Risberg, A.~Arendsen, D.~Davison, D.~Kopylenko, M.~Pollack, {\em et~al.}, ``The spring framework-reference documentation,'' {\em interface}, vol.~21, p.~27, 2004.

\bibitem{winikoff2005jack}
M.~Winikoff, ``Jack™ intelligent agents: an industrial strength platform,'' {\em Multi-agent programming: Languages, platforms and applications}, pp.~175--193, 2005.

\bibitem{broersen2001boid}
J.~Broersen, M.~Dastani, J.~Hulstijn, Z.~Huang, and L.~van~der Torre, ``The boid architecture: conflicts between beliefs, obligations, intentions and desires,'' in {\em Proceedings of the fifth international conference on Autonomous agents}, pp.~9--16, 2001.

\bibitem{agentframeworkNodeJS}
L.~Zhang, ``agentframework (2.0.1) [last access 13-11-2023],'' 2022.

\bibitem{spade}
S.~Palanca, Javi;~Alemany, ``Spade (3.3.2) [last access 13-11-2023],'' 2023.

\bibitem{varga2010omnet}
A.~Varga, ``Omnet++,'' in {\em Modeling and tools for network simulation}, pp.~35--59, Springer, 2010.

\bibitem{hinrichs2017distributed}
C.~Hinrichs and M.~Sonnenschein, ``A distributed combinatorial optimisation heuristic for the scheduling of energy resources represented by self-interested agents.,'' {\em IJBIC}, vol.~10, no.~2, pp.~69--78, 2017.

\bibitem{app9050923}
C.~Steinbrink, M.~Blank-Babazadeh, A.~El-Ama, S.~Holly, B.~Lüers, M.~Nebel-Wenner, R.~P. Ramírez~Acosta, T.~Raub, J.~S. Schwarz, S.~Stark, A.~Nieße, and S.~Lehnhoff, ``Cpes testing with mosaik: Co-simulation planning, execution and analysis,'' {\em Applied Sciences}, vol.~9, no.~5, 2019.

\bibitem{schrage2023multi}
R.~Schrage, P.~H. Tiemann, and A.~Nie{\ss}e, ``A multi-criteria metaheuristic algorithm for distributed optimization of electric energy storage,'' {\em ACM SIGENERGY Energy Informatics Review}, vol.~2, no.~4, pp.~44--59, 2023.

\bibitem{Schrage.2023}
R.~Schrage and A.~Nie{\ss}e, ``Influence of adaptive coupling points on coalition formation in multi-energy systems,'' {\em Applied Network Science}, vol.~8, no.~1, 2023.

\bibitem{tiemann2022operational}
P.~H. Tiemann, M.~Nebel-Wenner, S.~Holly, E.~Frost, A.~Jimenez~Martinez, and A.~Nie{\ss}e, ``Operational flexibility for multi-purpose usage of pooled battery storage systems,'' {\em Energy Informatics}, vol.~5, no.~1, pp.~1--13, 2022.

\bibitem{stark2021your}
S.~Stark, A.~Volkova, S.~Lehnhoff, and H.~de~Meer, ``Why your power system restoration does not work and what the {{ICT}} system can do about it,'' in {\em Proceedings of the twelfth ACM international conference on future energy systems}, pp.~269--273, 2021.

\bibitem{holly2021dynamic}
S.~Holly and A.~Nie{\ss}e, ``Dynamic communication topologies for distributed heuristics in energy system optimization algorithms,'' in {\em 2021 16th Conference on Computer Science and Intelligence Systems (FedCSIS)}, pp.~191--200, IEEE, 2021.

\end{thebibliography}

\end{document}